\documentclass[sigconf,screen,pbalance]{acmart}

\usepackage{fontawesome}
\usepackage{multirow}
\usepackage[inline,shortlabels]{enumitem}

\renewcommand\footnotetextcopyrightpermission[1]{}

\acmSubmissionID{1267}

\acmConference[DAC '26]{63rd ACM/IEEE Design Automation Conference}{July 26--29, 2026}{Long Beach, CA, USA}

\begin{document}

%%
%% The "title" command has an optional parameter,
%% allowing the author to define a "short title" to be used in page headers.
\title[MATCHA: Efficient Deployment of Deep Neural Networks on Multi-Accelerator Heterogeneous Edge SoCs]{MATCHA: Efficient Deployment of Deep Neural Networks\\ on Multi-Accelerator Heterogeneous Edge SoCs}

\author{
Enrico Russo\textsuperscript{$\dagger$},
Mohamed Amine Hamdi\textsuperscript{$\ddagger$},
Alessandro Ottaviano\textsuperscript{$\star$},
Francesco Conti\textsuperscript{$\S$}, \\
Angelo Garofalo\textsuperscript{$\star$$\S$},
Daniele Jahier Pagliari\textsuperscript{$\ddagger$},
Maurizio Palesi\textsuperscript{$\dagger$},
Luca Benini\textsuperscript{$\star$$\S$},
Alessio Burrello\textsuperscript{$\ddagger$}
}

\affiliation{
\textsuperscript{$\dagger$}University of Catania, Italy; 
\textsuperscript{$\ddagger$}Politecnico di Torino, Italy;
\textsuperscript{$\star$}ETH Zurich, Switzerland; 
\textsuperscript{$\S$}University of Bologna, Italy
\country{}
}

\thanks{
\raggedright
Corresponding authors: 
\texttt{enrico.russo@unict.it}, 
\texttt{mohamed.hamdi@polito.it}\break
\\

This work was accepted at the 63rd ACM/IEEE Design Automation Conference (DAC26)
}

%%
%% By default, the full list of authors will be used in the page
%% headers. Often, this list is too long, and will overlap
%% other information printed in the page headers. This command allows
%% the author to define a more concise list
%% of authors' names for this purpose.
\renewcommand{\shortauthors}{Russo et al.}
\renewcommand{\authors}{Enrico Russo, Mohamed Amine Hamdi, Alessandro Ottaviano, Francesco Conti, Angelo Garofalo, Daniele Jahier Pagliari, Maurizio Palesi, Luca Benini, Alessio Burrello}

%%
%% The abstract is a short summary of the work to be presented in the
%% article.
\begin{abstract}
  Deploying DNNs on System-on-Chips (SoC) with multiple heterogeneous acceleration engines is challenging, and the majority of deployment frameworks cannot fully exploit heterogeneity. We present MATCHA, a unified DNN deployment framework that generates highly concurrent schedules for parallel, heterogeneous accelerators and uses constraint programming to optimize L3/L2 memory allocation and scheduling. Using pattern matching, tiling, and mapping across individual HW units enables parallel execution and high accelerator utilization. On the MLPerf Tiny benchmark, using a SoC with two heterogeneous accelerators, MATCHA improves accelerator utilization and reduces inference latency by up to 35\% with respect to the the state-of-the-art MATCH compiler.
\end{abstract}

\maketitle

\section{Introduction}
\label{sec:introduction}
Deep Neural Networks (DNNs) are increasingly being deployed at the edge, where they must execute at high performance under tight latency, energy, and cost constraints~\cite{gillEdge2024}. To meet these competing objectives, modern edge systems-on-chip (SoCs) are becoming increasingly heterogeneous, integrating multiple specialized accelerators alongside general-purpose host CPUs~\cite{diana,carfield,dagli2022axonn}. 
Equipping a SoC with more than one kind of DNN accelerators enhances flexibility, as different architectural templates, e.g., Single-Instruction Multiple Data (SIMD)~\cite{flexpe}, vector units~\cite{spatz}, systolic or dataflow processors for General Matrix Mutiplication (GEMM)~\cite{diana,gemmini}, etc, provide different latency/energy efficiency trade-offs as a function of the accelerated operation(s) and corresponding tensors geometries~\cite{dagli2022axonn}. 
However, deployment frameworks for these SoCs often fail to fully exploit this flexibility, resulting in low hardware utilization, as they either offload entire DNNs to a single accelerator, or at most, map each DNN layer (or fused layer sequence) to the most suited compute unit, in a purely sequential fashion, controlled synchronously by the host~\cite{match,burrello2021dory,van2023htvm,scherer2024deeploy, park2024nest}. As a result, only one accelerator is active at any given time, while the others remain idle. 

In this paper, we examine two approaches for optimizing the utilization of available hardware resources. The first one leverages the structure of modern DNN computational graphs, which often include multiple independent branches (e.g., in ResNet-like \cite{resnet} networks or Attention blocks \cite{transformer}), that can be potentially offloaded in parallel to different accelerators. 
However, this approach is limited to multi-branch DNN models, and the optimization space is restricted to one-to-one layer to accelerator mappings, which are often insufficient to eliminate idleness entirely, as branches could be highly unbalanced in workload.

Thus, at a finer grain, we explore the parallelization opportunities provided by tiling DNN operators and assigning each tile to a different accelerator. While a trivial implementation of this offloading scheme is classic model parallelism, in which layer workloads are split evenly across homogeneous compute units~\cite{parallel_survey}, accelerators' heterogeneity in terms of operator support and performance makes that solution either sub-optimal or unfeasible, requiring a more general problem framing.

To fully support both these offloading schemes, we propose a novel DNN optimizing compiler that simultaneously explores layer-level and tile-level parallelization, targeting OS-less, multi-accelerator SoCs for edge inference. Our compiler is based on TVM~\cite{chen2018tvm} and on its extension MATCH for heterogeneous accelerators support~\cite{match}, but we completely redesign both the mapping optimization engine and the runtime to support concurrent, asynchronous execution on multiple compute units. Hence, we name it \textbf{MATCHA} (MATCH Asynchronous). Our main contributions are the following:
%\vspace{-0.2cm}
\begin{itemize}[leftmargin=*]
    \item We present a unified DNN deployment flow supporting concurrent offloading onto multiple accelerators, possibly with different internal architectures, thanks to a flexible model-based latency cost abstraction. To our knowledge, MATCHA is the first tool to support such kind of mapping for OS-less heterogeneous SoCs.
    \item We propose a constraint programming-based, tile-centric mapping framework to perform heterogeneous pattern matching and distribute work among accelerators, maximizing utilization.
    \item Through an experimental campaign on a complex heterogeneous SoC, equipped with a host CPU and two distinct accelerators, and on multiple full networks and layer blocks~\cite{banbury2021mlperf,resnet,resnext,transformer}, we show that MATCHA reduces end-to-end latency by up to 35\% with respect to the state-of-the-art MATCH compiler.
\end{itemize}

\section{Background and Related Work}
\label{sec:background}

\begin{table}

\definecolor{mygreen}{HTML}{15ba38}
\definecolor{myred}{HTML}{dd5b5b}
\definecolor{myblue}{HTML}{0298cc}
\def\sotasi{\textcolor{mygreen}{\faCheck}}
\def\sotano{\textcolor{myred}{\faClose}}
\def\sotani{\textcolor{orange}{\faMinus}}
\def\sotabo{\textcolor{myblue}{\faQuestion}}
\def\sotana{\textcolor{gray}{$\bullet$}}
\newcommand{\mychead}[1]{\rotatebox{90}{\parbox{1.6cm}{\linespread{0.8}\selectfont\textbf{#1}}}}

\centering
\caption{Comparison of state-of-the-art DNN deployment tools.}
\label{tab:sota}
\vspace{-0.65cm}
\resizebox{\columnwidth}{!}{
\begin{tabular}{@{}lccccccc@{}}
\textbf{Work}
  & \mychead{Open \\ Source}
  & \mychead{HSoC \\ Support}
  & \mychead{Operator \\ Extension}
  & \mychead{Device \\ Extension}
  & \mychead{Multi \\ ISA}
  & \mychead{Async. \\ Execution} 
  & \mychead{Pattern \\ Tiling} \\
\toprule

TFLite~\cite{david2021tensorflow} 
& \sotasi & \sotano & \sotasi & \sotasi & \sotana & \sotano & \sotana \\

TelaMalloc~\cite{maas2022telamalloc} 
& \sotano & \sotano & \sotani & \sotani & \sotana & \sotana & \sotano \\

Bolt~\cite{xing2022bolt} 
& \sotasi & \sotano & \sotasi & \sotano & \sotana & \sotani & \sotani \\

COMB~\cite{zheng2023memory} 
& \sotano & \sotasi & \sotasi & \sotasi & \sotano & \sotani & \sotano \\

HTVM~\cite{van2023htvm}
& \sotasi & \sotasi & \sotasi & \sotano & \sotano & \sotano & \sotano \\

DORY~\cite{burrello2021dory} 
& \sotasi & \sotano & \sotano & \sotano & \sotano & \sotano & \sotani \\

Deeploy~\cite{scherer2024deeploy} 
& \sotasi & \sotasi & \sotasi & \sotasi & \sotani & \sotani & \sotano \\

NEST-C~\cite{park2024nest} 
& \sotasi & \sotasi & \sotasi & \sotani & \sotani & \sotasi & \sotano \\

Map-and-Conquer~\cite{bouzidi2023map} 
& \sotano & \sotasi & \sotani & \sotani & \sotasi & \sotasi & \sotani \\

MATCH~\cite{match} 
& \sotasi & \sotasi & \sotasi & \sotasi & \sotano & \sotano & \sotano \\

\midrule
\textbf{Ours} 
& \sotasi & \sotasi & \sotani & \sotasi & \sotasi & \sotasi & \sotasi \\

\bottomrule \addlinespace[0.1cm] 
\multicolumn{8}{r}{ \footnotesize \sotasi\, Yes \,\,\, \sotano \, No \,\,\, \sotani \, Limited \,\,\, \sotana \, Not Applicable} \\
\end{tabular}
}
\vspace{-0.45cm}
\end{table}

Deploying deep neural networks efficiently on embedded and edge platforms has driven the design of a wide range of DNN accelerators and optimizing compilers. Despite the diversity of hardware and software stacks, most deployment frameworks follow a similar pipeline with four main stages:
\begin{enumerate*}[(i)]
    \item representing the DNN as a computation graph in an intermediate representation (IR) and applying graph-level transformations,
    \item partitioning the workload into subgraphs or tiles with an appropriate granularity,
    \item mapping each subgraph onto the available execution modules, and
    \item generating target-specific code.
\end{enumerate*}
The third step, commonly referred to as the \emph{mapping} phase, has been extensively studied for spatial and systolic DNN accelerators~\cite{symons2021loma,huang2021cosa,kao2020gamma,russo2023memory,parashar2019timeloop,yang2020interstellar}. Mapping decisions involve loop tiling (to fit large tensors into small local scratchpad memories), unrolling (to parallelize computation across the available processing elements), and ordering (to maximize data reuse and minimize data movement across the memory hierarchy).

Modern edge platforms increasingly adopt heterogeneous systems-on-chip (HSoCs), where a general-purpose host core is coupled with multiple specialized accelerators exposing different ISAs, data types, and memory organizations. Examples range from multi-processor SoCs (MPSoCs) that integrate CPU cores, GPUs, and NPUs on a shared DRAM (e.g., NVIDIA Xavier, Apple and Qualcomm SoCs)~\cite{dagli2022axonn}, to extreme edge-oriented SoCs that combine RISC-V control cores with tightly coupled accelerator clusters~\cite{diana,carfield}. In this work we target edge HSoCs that offer several accelerators and processor clusters connected through a software-managed, multi-level scratchpad memory hierarchy. 

Table~\ref{tab:sota} compares representative DNN deployment frameworks for these systems along key dimensions: open-source availability, explicit support for heterogeneous SoCs, extensibility of operator and device backends, support for multi-ISA systems, asynchronous execution capability, and tile-level inter-device parallelization.
TFLite~\cite{david2021tensorflow} provides portability and pluggable kernels but offers limited control over mapping and tiling.
DORY~\cite{burrello2021dory} optimizes memory allocation for single accelerators using constraint programming, while TelaMalloc~\cite{maas2022telamalloc} and Bolt~\cite{xing2022bolt} target specific subproblems such as memory allocation or GPU tuning without full heterogeneous support.
More recent tools, including Deeploy~\cite{scherer2024deeploy}, HTVM~\cite{van2023htvm}, NEST-C~\cite{park2024nest}, COMB~\cite{zheng2023memory}, and Map-and-Conquer~\cite{bouzidi2023map}, extend compiler frameworks to heterogeneous SoCs, but differ in their granularity and scheduling scope.
Deeploy and HTVM combine memory allocation and tiling, yet primarily target coarse-grained offloading and \emph{sequential} execution.
COMB focuses on mapping multiple DNN models simultaneously to leverage graph-level parallelism and does not explore asynchronous tile-centric execution. Map-and-Conquer focuses on collaborative execution on \emph{OS-equipped} MPSoCs, which removes the need for explicit memory management and simplifies layer tiling. Furthermore, it considers a trivial workload offloading based on accelerators' peak performance. MATCH~\cite{match}, the closest to our work, extends TVM with hardware-aware design-space exploration for HSoCs to assign layers or fused layer groups (patterns) to accelerators. Nonetheless, MATCH still maps at layer/pattern granularity and executes kernels sequentially.

\section{Methodology}
\label{sec:methodology}

\begin{figure}
    \centering
    \includegraphics[width=\linewidth]{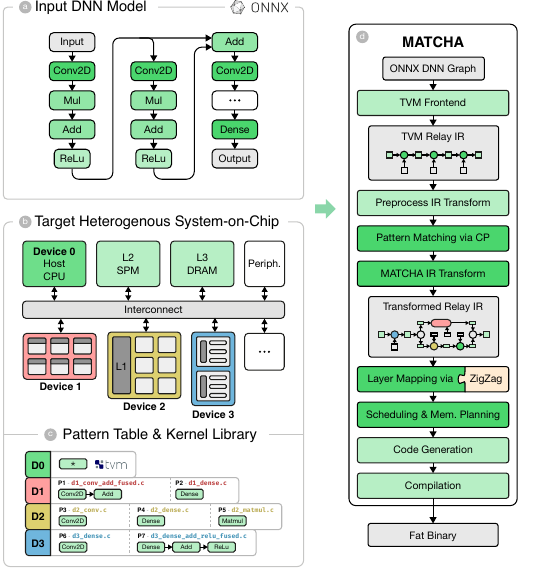}
    \caption{MATCHA deployment framework: inputs (left) and pipeline stages (right).}
    \label{fig:overview}
    
\end{figure}

In the following, we refer to each execution module (host or accelerator) of the HSoC, capable of running a DNN kernel, as a \textit{device}. The overall MATCHA\footnote{MATCHA is available open-source at: \url{https://github.com/eml-eda/match}.} pipeline is depicted in Fig.~\ref{fig:overview}. MATCHA takes as input: a DNN model in ONNX format (Fig.~\ref{fig:overview}a), a description of the target HSoC (Fig.~\ref{fig:overview}b), including available devices and memory capacities, and a catalogue of supported layer patterns together with the kernels provided by each device (Fig.~\ref{fig:overview}c)

The pipeline begins by importing the ONNX model into the TVM~\cite{chen2018tvm} Relay intermediate representation (IR), a directed graph in which nodes represent tensors or primitive operators and edges denote data flow. A pre-processing pass applies useful graph transformations to the Relay IR, such as constant folding or dead nodes removal.

In MATCHA, the pattern-matching optimization is then performed by a constraint-programming (CP) optimizer that maps each layer tile to the most appropriate computing device; details are discussed in Section~\ref{sec:pattern}. Based on the optimizer's output, MATCHA rewrites the IR: operators are split and tiled according to the chosen tiling strategy, fused kernel supernodes, and auxiliary operators (e.g., tensor slicing and concatenation) are added to the IR, and the graph is partitioned to map operators to devices. The transformed graph is subsequently subject to scheduling and memory planning (Section~\ref{sec:scheduling}). In the scheduling phase, the mapped tiles to each device are scheduled to meet local multi-level memory requirements: loop tiling, unrolling, and ordering are optimized using the ZigZag DNN layer mapper~\cite{mei2021zigzag}. Finally, as described in Section~\ref{sec:code_generation}, MATCHA emits C code for the host and accelerators and produces a multi-architecture binary image using target-specific compilation tools.

\subsection{Pattern Matching and Layer Tiling}
\label{sec:pattern}

\begin{figure}[!t]
    \centering
    \includegraphics[width=\linewidth]{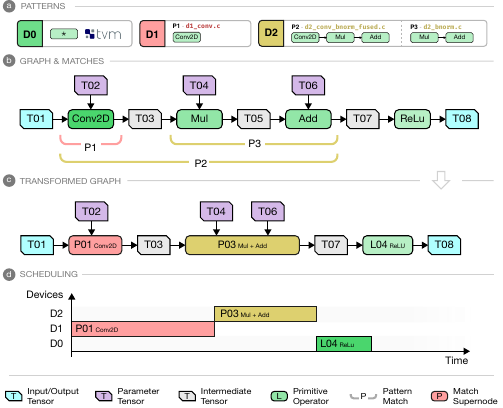}
    \caption{Conventional pattern matching in a heterogeneous system resulting in sequential execution across devices.}
    \label{fig:matching_match}
\end{figure}

We model the Relay IR as a directed graph $\mathcal{G}_{\mathrm{IR}}=(\mathcal{V},\mathcal{E})$, where nodes (vertices) $\mathcal{V}$ are either tensors or primitive operators, and edges~$\mathcal{E}$ represent data dependencies. An operator pattern is modeled as a path (chain) graph  $ p=(\mathcal{V}_{p},\mathcal{E}_{p}) $ of length $ \ell_{p} $ with node-level constraints (e.g., operator types, tensor shapes and layouts, data type and precision). In conventional deployment frameworks, pattern matching seeks injective graph homomorphisms $ h:\mathcal{V}_{p}\to\mathcal{V} $ that satisfy these constraints; then, each selected match replaces the corresponding subgraph with a fused supernode. Figure~\ref{fig:matching_match}b illustrates an IR example and highlights possible pattern matches; Figure~\ref{fig:matching_match}c shows the resulting IR after matching decisions and graph transformation. As shown in Figure~\ref{fig:matching_match}d, when the original graph has no branches, this approach leads to sequential execution across devices as each fused supernode is mapped to a single accelerator.

MATCHA generalizes this paradigm to maximize utilization on heterogeneous SoCs by introducing a tile-centric formulation that (i) allows pattern matches to cover integer numbers of tiles of underlying operators, and (ii) enables asynchronous and parallel execution of tiles across devices. The optimizer
jointly selects pattern matches, tile allocations, and device mappings to minimize end-to-end latency.

\begin{figure}[!t]
    \centering
    \includegraphics[width=\linewidth]{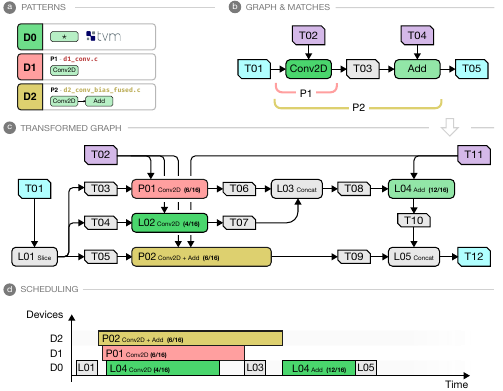}
    \caption{MATCHA's tile-centric pattern matching and tiling with asynchronous execution across heterogeneous devices.}
    \label{fig:matching_ours}
\end{figure}

Each IR operator $v\in\mathcal{V}$ is partitioned into an integer number $ T_v $ of tiles (e.g., feature map rows for convolutional layer or output neurons in dense layers). Let $ \mathrm{Ops}_v $ denote the total arithmetic operation count of operator $ v $. The available kernel pattern library is denoted by $\mathcal{P}$ (Fig.~\ref{fig:matching_ours}a) and each pattern \(p\in\mathcal{P}\) is associated with device $d_p$, an efficiency factor $\eta_{p}\in(0,1]$, and a fixed per-invocation time overhead \(\delta_{p}\). In addition to accelerator-specific kernels, MATCHA always includes a \textit{wildcard} pattern that covers any individual operator with a TVM-generated kernel ensuring that unmatched tiles can be executed by the host.

Each match $m$ of the pattern $p$ defines a mapping \mbox{$h_{m}:\mathcal{V}_{p}\to\mathcal{V}$} that identifies which IR nodes are involved. For each match, MATCHA introduces a nonnegative integer decision variable \(t_{p,m}\) that represents the number of tiles assigned to it. The same \(t_{p,m}\) applies to every IR operator \(v\) in the image of \(h_m\). Tile conservation across all instantiated matches is enforced as:
\begin{equation}
\label{eq:tile-conservation}
\sum_{p\in\mathcal{P}}\ \sum_{m\in\mathcal{M}_p} \mathbb{I}_{v,p,m}\, t_{p,m} \;=\; T_v
\qquad\forall v\in\mathcal{V},
\end{equation}
where $\mathbb{I}_{v,p,m} = 1$ if $v\in h_m(\mathcal{V}_{p})$ and $0$ otherwise and $\mathcal{M}_p$ is the set of matches of $p$.

The optimizer estimates the latency of each instantiated match using a lightweight analytical model. For a fused match~$m$ of pattern~$p$ on device~$d_p$ to which $t_{p,m}$ tiles are assigned, latency is calculated as:
\begin{equation}
\label{eq:match-latency}
L_{p,m}(t_{p,m})
\;=\; t_{p,m}\cdot\left(\sum_{u\in\mathcal{V}_{p}} \frac{\mathrm{Ops}_{\,h_{m}(u)}}{T_{\,h_{m}(u)}}\right)\frac{\alpha_{d_p}}{\eta_{p}}
\;+\; \delta_{p},
\end{equation}
where \(\alpha_d\) denotes the device speed parameter (time per arithmetic operation, i.e., the inverse of peak operations per time unit). The inner sum yields the per-tile arithmetic work of the fused pattern under match \(m\). Multiplying by \(t_{p,m}\) produces the total arithmetic work executed by the match supernode. The device speed, kernel efficiency, and fixed overhead translate assigned work into execution time. A more advanced model that takes into account tile splitting overhead and specific operator geometry is currently not used, but can be explored in future work.

Because the tile variables \(t_{p,m}\) simultaneously determine which supernodes are instantiated, how much work they perform, and what work remains for other devices, MATCHA frames joint pattern-matching, device-allocation, and \emph{platform-scheduling} (as tiling-to-device assignment) as a constrained optimization problem whose objective is the overall makespan. The linear dependence of the latency formulas on the tile variables makes the cost evaluation tractable within CP. Additional constraints encode device memory capacities and concurrency feasibility. Fig.~\ref{fig:matching_ours}c shows a representative pattern matching solution: assume a \textit{Conv2D} kernel on device~1 and a fused \textit{Conv2D+Add} kernel on device~2; both \textit{Conv2D} and the subsequent \textit{Add} are partitioned into \(T=16\) tiles. Two match instances of the fused pattern are selected: one match (pattern \textit{Conv2D}) assigned 6 tiles to device~1, and one match (pattern \textit{Conv2D+Add}) assigned 6 tiles to device~2; the remaining 4 \textit{Conv2D} tiles and the remaining 12 \textit{Add} tiles are handled by host.

After the CP optimizer selects a tiling and matching configuration, MATCHA instantiates the corresponding fused supernodes and helper operators in the IR (e.g., slice, concat, etc.) and proceeds to a final \emph{device-specific scheduling} and memory-planning stage (Section~\ref{sec:scheduling}). Notice that layer-device assignment and asynchronous execution becomes a corner case of this optimization, in which each tile of the layers is assigned to the same device.
\vspace{-0.2cm}
\subsection{Mapping, Scheduling and Memory Planning}
\label{sec:scheduling}
This stage refines the device-specific scheduling, computes tensor placements and lifetimes, and produces a temporally ordered execution plan that respects data dependencies and memory constraints.

\begin{figure}
    \centering
    \includegraphics[width=0.9\linewidth]{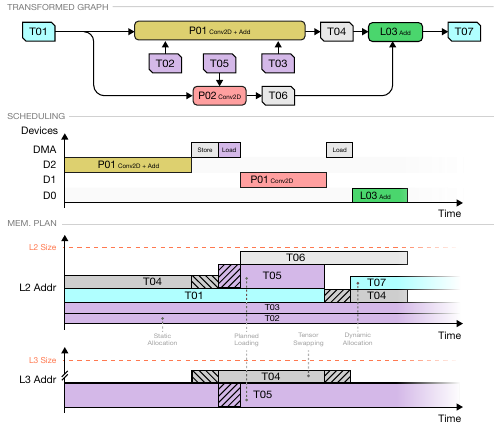}
    \vspace{-0.2cm}
    \caption{Scheduling and memory plan example with different types of tensor allocation.}
    \label{fig:memplan}
    \vspace{-0.4cm}
\end{figure}

The optimizer models the joint scheduling and memory-allocation problem as a two-dimensional bin-packing instance (time \(\times\) address) on the on-chip main memory (L2) and the larger off-chip memory (L3). Tensor lifetimes induce temporal occupancy intervals; the planner may choose among allocation strategies:
(i) static allocation: assign a persistent address in L2 and keep the tensors resident throughout execution (always alive); (ii) dynamic allocation with optional swapping: evict intermediate variable tensors to L3 after production and reload them later when required; 
(iii) planned loading: load a large parameter tensor from L3 on demand.
When swapping or planned loading is selected, data movement between L2 and L3 is modeled and accounted for in the makespan: transfers are performed by DMA engines and are serialized in the current model (i.e., DMA transfers do not overlap with computation). Fig.~\ref{fig:memplan} shows a simple memory plan and scheduling example where each node is assigned to a different device: even if enabled, asynchronous execution can not happen, given that the optimizer forces a sequential scheduling to fit the constrained memory (T04, T05, and T06 do not need to be stored simultaneously).

At this stage, the computation latency estimates are refined to reflect lower-level mapping decisions. Operators assigned to an accelerator frequently cannot place all working data in the accelerator's local scratchpad (L1); therefore additional tiling between L1 and L2 (e.g., loop unrolling, ordering) is applied and evaluated using a device-level mapper and cost model. MATCHA uses the ZigZag LOMA \cite{mei2021zigzag,symons2021loma} mapper together with its cost model to select L1/L2 tilings and memory access schedules; the resulting per-node latency estimates (including any necessary data movement) are considered to perform the global schedule.

Again, CP optimization is used. It enforces device memory capacity constraints, concurrency limits (e.g., each device can run one kernel at a time), and data-dependency precedence, and minimizes the makespan subject to these constraints. The optimizer outputs (i) an execution schedule that orders kernel invocations and DMA operations, (ii) a memory plan that specifies address assignments and swap points, together with (iii) a detailed layer-to-device mapping along with the L1/L2 tiling choices obtained through ZigZag.

These artifacts are then translated into low-level code generation primitives and passed to the backend (see Section~\ref{sec:code_generation}) to produce the binaries. The final result of this stage is an executable schedule and memory plan that together enable asynchronous, parallel execution of the DNN across the heterogeneous SoC.

\subsection{Multi-device Code Generation}
\label{sec:code_generation}

The code-generation stage consumes the transformed Relay IR together with the execution schedule, the planned tensor addresses, and the per-layer tiling solution (loop tiling, unrolling and ordering). MATCHA uses a collection of Mako templates~\cite{mako} to synthesize C implementations for each executable node (layer) in the transformed graph. In the HSoC platform specification, the user additionally provides a small set of platform-level C APIs that the generator may call for device-specific services such as device initialization, DMA submission and completion, inter-core synchronization, and allocation/management of scratchpad buffers.

MATCHA supports targets that expose symmetric multiprocessing (SMP) semantics. When available, platform APIs are used to obtain core identifiers and to implement intra-device synchronization and work distribution across cores.
Alongside per-node kernel code, the generator emits a graph runtime that executes on the host. The host-side runtime exposes a single entry point that starts inference for a given input tensor; on invocation it initializes the planned memory pool, writes the tensor base addresses according to the memory plan, and drives execution by issuing kernel invocations and DMA operations in the temporal order produced by the optimizer. The host runtime is responsible for orchestrating cross-device execution, honoring data dependencies, and invoking any host-resident layer implementations.

MATCHA is designed for heterogeneous SoCs in which accelerators may implement a different ISA from the host and therefore require a distinct compiler toolchain and runtime. To accommodate such platforms (including bare-metal edge SoCs that lack a full operating system), MATCHA can emit a lightweight device-side runtime for each accelerator. The device runtime implements a simple dispatch loop: it waits for a task descriptor from the host together with input tensor addresses, executes the corresponding layer kernel with the locally mapped tiling and memory plan, and then signals completion back to the host. This separation permits each device image to be compiled with the toolchain appropriate to its ISA and execution environment.

The host and devices communicate and synchronize using one of two supported methods: (i) polling, in which the host and/or device busy-wait on shared status flags or memory locations, or (ii) event/interrupt-based notification, in which the host signals a device via an interrupt or event and the device signals completion likewise. The event/interrupt mechanism enables low-overhead, \emph{asynchronous overlap of each device}; the host runtime uses the optimizer's schedule together with online device availability to correctly dispatch tasks and execute host-resident layers.

After code generation, host and device sources are compiled with the appropriate compilers and toolchains. The compilation outputs are packaged into a single multi-architecture binary image. At startup the host loads and initializes each device runtime; once initialization completes the devices are ready to accept tasks according to MATCHA's generated schedule.

\section{Experimental Results}
\label{sec:results}

\begin{figure}
    \centering
    \includegraphics[width=0.9\linewidth]{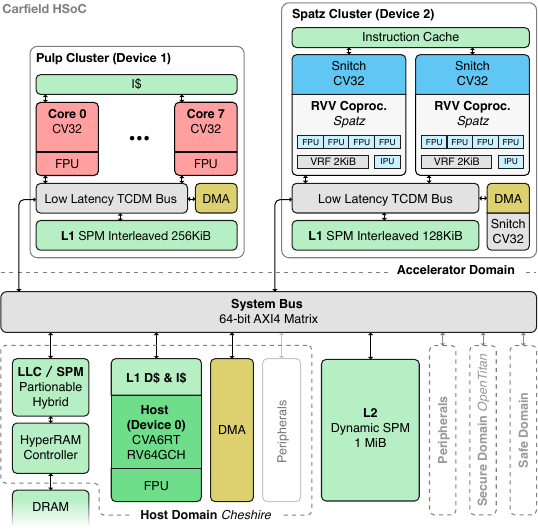}
    \caption{Use case Carfield HSoC considered for evaluation.}
    \label{fig:carfield}
\end{figure}

We implement MATCHA in Python and rely on Apache TVM~\cite{chen2018tvm} for host-side kernels, the ZigZag LOMA~\cite{mei2021zigzag,symons2021loma} mapper for device-level tiling and mapping, Mako for template-based code generation, and OR-Tools for constraint programming solvers. To evaluate MATCHA, we use the open-source Carfield heterogeneous SoC~\cite{carfield} as our target platform. The design was deployed on a Xilinx VCU118 FPGA and experiments were conducted at 50 MHz clock frequency.
\begin{table*}

\centering
\caption{MLPerf Benchmark results. MATCH employs always the best accelerator for each layer (Spatz / Pulp).}
\label{tab:mlperf}
\vspace{-0.3cm}
\resizebox{\textwidth}{!}{
\begin{tabular}{@{}lrrccccccccccccc@{}}
 &  &     
& \multicolumn{3}{c}{TVM}
& \multicolumn{3}{c}{MATCH}
& \multicolumn{3}{c}{MATCHA (No Tiling)}
& \multicolumn{3}{c}{MATCHA (Ours)} \\
\cmidrule(l){4-6}
\cmidrule(l){7-9}
\cmidrule(l){10-12}
\cmidrule(l){13-15}
Model & MACs & Params 
& Cycles & Runtime & FLOPS 
& Cycles & Runtime & FLOPS
& Cycles & Runtime & FLOPS
& Cycles & Runtime & FLOPS \\
\toprule

AutoEncoder & 0.27M & 268k & 
5.03M & 100.2 ms & 5.36M & 
1.01M & 20.1 ms & 26.7M & 
1.01M & 20.1 ms & 26.7M & 
0.673M & \textbf{13.4 ms} & 40.1M \\

DS-CNN & 2.8M & 22.6k & 
30.3M & 604.6 ms & 9.29M &
6.58M & \textbf{131.1 ms} & 42.9M &
6.58M & \textbf{131.1 ms} & 42.9M &
6.58M & \textbf{131.1 ms} & 42.9M \\

MobileNet & 7.9M & 210k & 
157.4M & 3137.8 ms & 5.07M & 
24.4M & \textbf{486.7 ms} & 32.7M & 
24.4M & \textbf{486.7 ms} & 32.7M & 
24.4M & \textbf{486.7 ms} & 32.7M \\

ResNet18 & 12.8M & 78k & 
200.2M & 3991.7 ms & 6.43M &
22.9M & 456.6 ms & 56.2M & 
19.9M & 395.9 ms & 64.9M & 
16.3M & \textbf{325.1 ms} & 79.0M \\

\bottomrule
\end{tabular}
}
\end{table*}

Carfield is a time-predictable, HSoC intended for mixed-criticality, AI-enhanced sensor-processing and control workloads (e.g., automotive and space applications). The configuration used in our experiments is illustrated in Fig.~\ref{fig:carfield}. The SoC comprises a host domain based on the Cheshire platform~\cite{cheshire}, which integrates a dual-core RV64GCH 64-bit RISC-V CPU, and an accelerator domain containing two heterogeneous accelerator clusters. Conventional peripherals include a system DMA that can be used for L2-L3 transfers; the system interconnect is a 64-bit AXI4 bus. A 1\,MiB on-chip, dynamically configurable L2 scratchpad is shared among domains and is exposed to the interconnect with a 128-bit-per-cycle data path.

The accelerator domain contains two accelerators. The first is a PULP cluster~\cite{rogenmoser2025hybrid} composed of eight 32-bit RISC-V (RI5CY) cores with floating-point ISA extensions. 
The cluster provides a 256\,KiB L1 scratchpad and a dedicated DMA engine for L2–L1 transfers.

The second accelerator is a Spatz cluster~\cite{spatz}, consisting of two compact scalar RISC-V cores that control two RISC-V Vector Units (RVVUs                               ). Each RVVU implements RVZve64d semantics with a vector length VLEN = 512 bits and supports data formats from FP8 up to FP64, bfloat16, integer types, and mixed-precision primitives (including sum-of-dot-product, sdotp). The Spatz cluster includes a 128\,KiB L1 scratchpad and its own DMA engine for L2–L1 transfers.

The Carfield design also exposes a Platform-Level Interrupt Controller (PLIC) for centralized host interrupt handling and per-device interrupt-triggering mailbox units that interface the host to each accelerator cluster and other devices. MATCHA can take advantage of these asynchronous mailbox notifications to implement low-latency task dispatch and completion signaling, enabling parallelism and overlap between host and accelerator execution.

This heterogeneous HW configuration with diverse cores and accelerators, shared on-chip scratchpad, and explicit DMA engines, provides a representative testbed for MATCHA's tile-centric pattern matching, device allocation, and schedule-aware memory planning. 

We use MATCH \cite{match} as the state-of-the-art baseline because it can be easily extended to support the target platform, allowing a fair comparison. No other compilers from Table \ref{tab:sota} support the Carfield architecture. All the following experiments employ the same set of patterns and kernels across the evaluated toolchains, considering FP16 data precision.

\paragraph{Microbenchmarks}

\begin{figure}
    \centering
    \includegraphics[width=\linewidth]{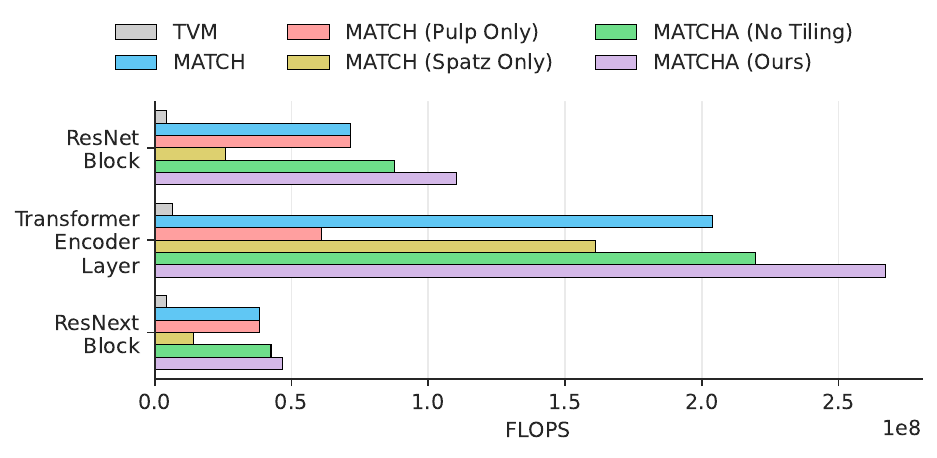}
    \caption{FLOPS comparison for DNN benchmark blocks.}
    \label{fig:exp_blocks}
\end{figure}

Figure~\ref{fig:exp_blocks} reports floating-point operations per second (FLOPS) for three representative DNN building blocks evaluated on the Carfield HSoC: the first residual block of ResNet-50 (three convolutional layers), the first block of ResNeXt-50~\cite{resnext}, and a Transformer encoder layer (multi-head attention, feed-forward, and normalization) with hidden size 128. 
Compared to the TVM host-only baseline, MATCHA achieves speedups ranging from $11.04\times$ to $40.34\times$ across these blocks. 
When only enabling asynchronous execution and leveraging graph-level parallelism, MATCHA yields latency reductions of $18.22\%$, $7.21\%$, and $9.47\%$ for the ResNet-50, Transformer, and ResNeXt-50 blocks, respectively, relative to MATCH allocating each layer on the best-performing accelerator. 
Enabling MATCHA’s tile-centric optimization further improves load balancing across devices, reducing latency by $35.02\%$ on the ResNet-50 block, $17.55\%$ on the ResNeXt-50 block, and $23.65\%$ on the Transformer encoder layer compared to MATCH.
Overall, the achievable speedups are bounded by the performance imbalance between heterogeneous accelerators and by the fraction of operators in each block that are not supported and must execute on the host (with two homogeneous accelerators and with all operators offloaded, the maximum speedup compared to using a single one would be  50\%). However, they demonstrate that tile-centric asynchronous execution is crucial to maximize HSoCs HW utilization.

\begin{figure}
    \centering
    \includegraphics[width=\linewidth]{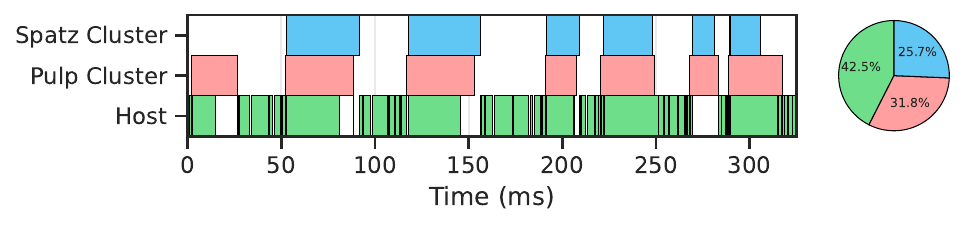}
    \caption{ResNet inference profiling timeline (left) and the execution time breakdown (right) across different devices.}
    \label{fig:exp_resnet_timeline}
\end{figure}

\paragraph{MlPerf}

Table~\ref{tab:mlperf} reports inference cycles, latencies and FLOPS of MLPerf Tiny~\cite{banbury2021mlperf} benchmark models for TVM, MATCH (sequential execution choosing the fastest device for each operator), MATCHA with pattern tiling disabled (asynchronous layer offloading only), and MATCHA with pattern tiling enabled. In these experiments, patterns including convolutional layers are tiled along the rows of the output feature map; this tiling requires inserting input slicing operations and output concatenations, which introduce additional runtime overhead. For compact networks dominated by depthwise convolutions (e.g., DS-CNN and MobileNet) we observe no latency improvement relative to the baseline. Depthwise convolutions exhibit low arithmetic intensity with respect to traditional convolutional layers; therefore, the overheads of slicing and concatenation outweigh any computational benefit from tiling. This result highlights the importance of optimizing slice/concat primitives and automatically identifying the optimal tiling dimension, as discussed in Sec.~\ref{sec:conclusion}. 
For ResNet-18 trained on CIFAR-10, MATCHA already benefits from asynchronous cross-device execution of residual branches and yields a $13.3\%$ latency reduction. Enabling MATCHA's intra-layer tiling to increase device utilization produces a larger reduction, reaching $28.8\%$ relative to MATCH. The AutoEncoder model, commonly used for anomaly detection, consists of a chain of fully connected layers, which offer little opportunity for graph-level parallel execution. Nonetheless, MATCHA's pattern-tiling capability achieves a $33.3\%$ latency reduction versus MATCH. We tile fully connected layers across the output-neuron dimension; since the corresponding weight tiling can be folded into the offline weight layout, this strategy incurs essentially zero runtime overhead. Overall, MATCHA outperforms the TVM host-only compilation by factors ranging from $4.61\times$ to $12.28\times$ across the evaluated networks. Fig.~\ref{fig:exp_resnet_timeline} shows the ResNet inference profiling timeline and the execution time breakdown showing a balanced workload distribution; concurrent execution is interleaved by unmatched activation operators 
and helper operators running exclusively on the host.

\section{Conclusions}
\label{sec:conclusion}

We presented MATCHA, a tile-centric deployment framework that employs constraint programming to perform pattern matching, device allocation, scheduling, and memory planning, enabling asynchronous DNN inference on heterogeneous SoCs. On the representative Carfield SoC, MATCHA yields end-to-end speedups versus prior SoA methods of 35\%, demonstrating that tile-aware pattern matching and asynchronous execution are effective approaches to increase utilization in HSoCs.

While already providing latency reduction, MATCHA still presents some limitations which will be addressed in future work to unlock even higher utilization: (i) the flow is split between the coarse pattern matcher and the detailed scheduling per accelerator to reduce the overall search space. However, jointly optimizing pattern matching, per-device layer mapping, and low-level tiling in a single cost model (potentially with hardware-in-the-loop parameter calibration) could unlock further performance improvement. (ii) Helper-operator overheads (slice/concat) are not accurately modeled in latency models. Moreover, these operators can be eliminated by view-based tiling or carefully planned contiguous tensor address placements. Lastly, the assessment of the energy improvement of asynchronous execution is not evaluated.

\begin{acks}
This work was supported by European Commission through the Chips Joint Undertaking under grant agreement number 101139790 (ECS4DRES). Part of this work was carried out while Enrico Russo was visiting the Integrated Systems Laboratory (IIS), ETH Zurich.
\end{acks}

\bibliographystyle{ACM-Reference-Format}
\bibliography{references}

\end{document}